\documentclass[twocolumn]{revtex4}
\usepackage{graphicx}

\begin{document}
\title{Folded Graphene Nanoribbons with Single and Double Closed Edges}

\author{Nam B. Le}
  \affiliation{Department of Physics, University of South Florida, 4202 E. Fowler Avenue, Tampa, FL 33620}
  \affiliation{Institute of Engineering Physics, Hanoi University of Science and Technology, 1 Dai Co Viet, Hanoi, Vietnam}

\author{Lilia M. Woods}
  \affiliation{Department of Physics, University of South Florida, 4202 E. Fowler Avenue, Tampa, FL 33620}

\begin{abstract}
Graphene nanoribbon folds with single and double closed edges are studied using density functional theory methods. Van der Waals dispersive interactions are included via semi-empirical pairwise optimized potential. The geometrical phases of the single and double folded ribbons are obtained. The electronic structure in terms of energy needed for the folding process, van der Waals contribution, energy band gaps, and bandstructures are also calculated. The results are interpreted in terms of peculiarities of the structures and dispersion interactions. It is shown that significant modifications in the electronic structure can be achieved as a result of folding.
\end{abstract}

\pacs{61.48.Gh,81.05.ue,73.22.Pr,68.65.Pq}
\maketitle

The recent isolation of a single graphene layer has generated much scientific interest\cite{Science306-666,PNAS102-10451}. With its extraordinary electrical, optical, and mechanical properties, this atomically thick two-dimensional material is highly expected to be the material of novel technological devices, especially for high-speed electronics \cite{NL9-4474,Science372-662,Nature467-305,Nature474-64}. Graphene nanoribbons (GNRs), which are quasi-one dimensional strips of graphene, have also been synthesized \cite{PRL98-206805,Nature458-872,NL8-2773}. GNRs have their own unique characteristics. It turns out that the types of edges and the magnitude of the width largely determine the metallic or semiconducting nature of GNRs. Magnetic states are also found to exist in zigzag-terminated nanoribbons \cite{PhysicaE40-228,PRL97-216803,JPSJ65-1920}.

Graphene and GNRs can be easily folded in the out-of-plane direction. Such wrapping changes their functionalities and properties. Folded GNRs  (FGNRs) can be viewed as two flat sheets connected by a curved edge. They are also referred to as fractional nanotubes because the curved edge has a nanotube-like structure. Graphene folds have been observed experimentally \cite{ACSNN4-5480}, and they can also be achieved by mechanical stimulations \cite{PRL102-015501,PRL104-166805} or by high temperature annealing \cite{SurfSci407-1}. Recently, a variety of different forms of graphene pleats have also been demonstrated \cite{PRB83-245433,EPJB64-249}. The majority of natural graphite occurs in the Bernal (AB) stacking sequence, and only a small portion appears in the rhombohedral ABC form \cite{ProcRSocLondonSerA181-101}. However, scanning tunneling microscopy studies have shown that FGNRs can occur not only in $AB$, but also in $AA$ or other patterns \cite{SurfSci407-1,PRB83-245433}. Thus folding can offer the possibility to generate other registry dependent configurations.   

The existence of interlayer coupling together with the edges at the open ends and bending energy at the closed edge influence the stability, and electronic and transport characteristics in a profound way. Recent studies have developed a coarse-grained model showing that the balance between the bending and adhesion determines self-folding \cite{APL95-123121}. Nearest-neighbor tight binding reports have calculated the conductance of FGNRs showing that the electronic structure can be changed significantly by assuming different values of the adhesion between the flat portions \cite{JAP106-103714}. Electronic structure calculations have revealed that FGNRs have permanent electric dipoles \cite{PRB80-165407}.

In this work we investigate FGNR structures with single and double closed edges. The open sides of the studied structures have H saturated armchair ends. The folded configurations are found to have a variety of unusual stacking patterns. Our {\it ab initio} calculations reveal the evolution of the geometries as a function of the width of the ribbons. Particular emphasis is placed on the role of the van der Waals (vdW) dispersion and its role in terms of the stability of the various formations. The electronic structure is also obtained showing that single and double folds can have much different properties as compared to those of their planar counterparts.

{\bf Method and Structures.} Our calculations are performed using self-consistent density functional theory (DFT) implemented in the Vienna Ab-Initio Simulation Package (VASP) \cite{vasp}. The code uses a plane-wave basis set and periodic boundary conditions. Here we utilize PAW-PBE potentials to account for the core electrons \cite{PRB50-17953,PRB59-1758}. For all studied systems, $(1\times1\times7)$ Monkhorst-Pack k-grid sampling of the Brillouin zone was taken for the self-consistent calculations with an energy cutoff of $430\,eV$. All structures are allowed to relax with $10^{-5}$ eV total energy and $0.02\,eV/\mbox{\it\AA}$ force convergence criteria. Each supercell is contructed so that the distance between the nearest neighbors in each directon is about 10 {\it\AA}.

The interlayer coupling is largely determined by the long ranged van der Waals (vdW) interaction, which is crucial for the formation and stability of the folds. Here it is taken into account via the recently developed DFT-D2 method implemented in VASP \cite{JPCA114-11814}. This is a pragmatic approach that calculates the vdW interactions originating from the fluctuating charge distributions via a semi-empirical correction to the conventional Kohn-Sham energy. This correction is described with an optimized pairwise force field weighted with a damping function. Full parametrization for the entering coefficients for different elements was achieved in Refs. \cite{JCC25-1463,JCC27-1787}.

\begin{figure}[t]
  \includegraphics[scale=0.2]{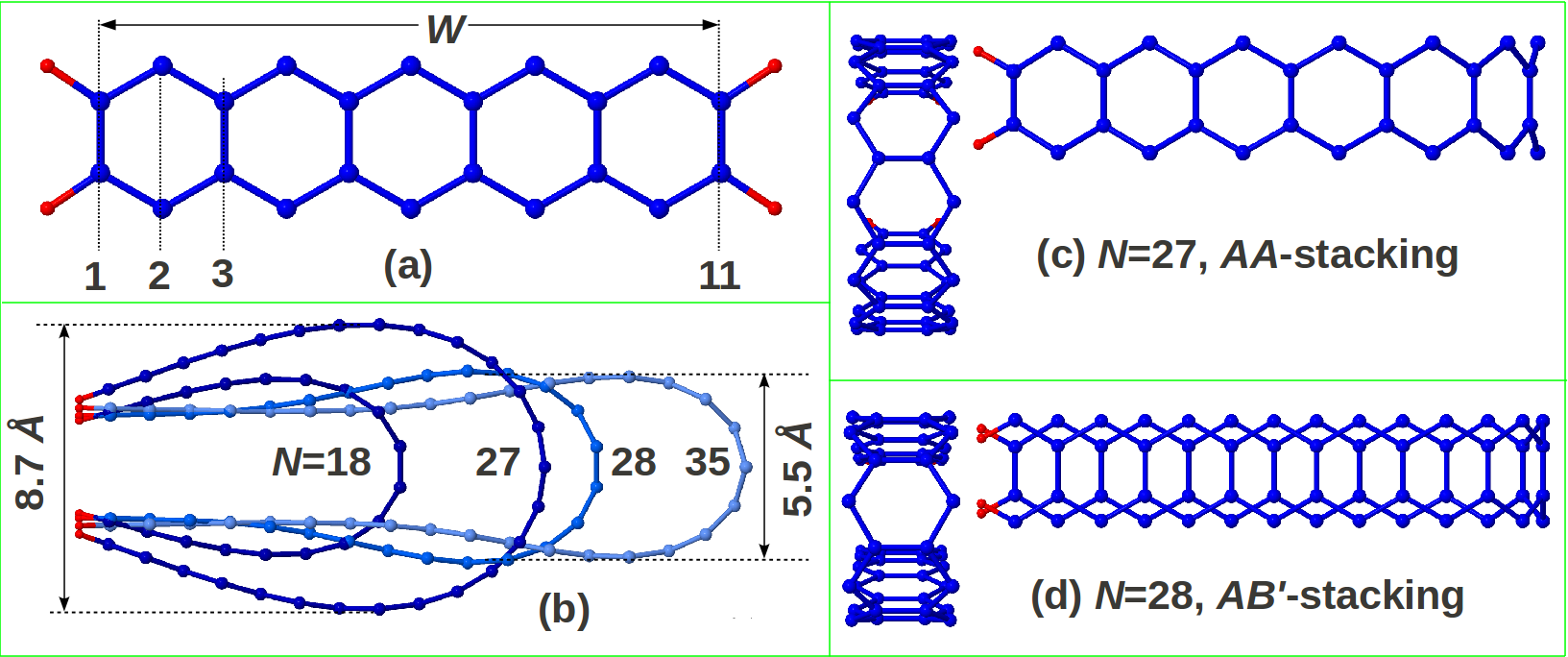}
  \caption{(Color Online) (a) An armchair GNR with H-saturated armchair ends for $N=11$ Carbon lines. The width of the ribbon is $W$; (b) Front view of the relaxed single FGNRs with $N=$ 18, 27, 28, and 35 Carbon lines. Side and top views (from left to right in each panel) of the relaxed structures for (c) $N=27$ - $AA$ stacking; and (d) $N=28$ - $AB'$ stacking.}
  \label{fig1}
\end{figure}

We study armchair edged GNRs, which are nonmagnetic and of semiconducting nature with energy band gaps decreasing as the width of the ribbon becomes larger. The ends of the ribbons are saturated by H bonds in order to avoid chemical self-interaction. Each GNR can be represented via the number of C atomic lines $N$ along its axis, as shown in Fig.\,\ref{fig1}a. A single FGNR is prepared by bending the flat ribbon onto itself creating a symmetric structure with two parallel graphene portions lying on top of each other. A double FGNR is done in a similar manner but folding the GNR twice with three graphene portions on top of each other.

{\bf Results and Discussion.} Each fold is relaxed within the above specified criteria. The obtained structures of some of the single folded ribbons are shown in Fig.\,\ref{fig1}b. We find that when $17\leq N \leq 27$, the single fold takes a shape that looks like a racket without a handle. The distance between the furthest points on the curved side increases as the width of the ribbon becomes larger with a maximum value of $8.7$ {\it\AA}\, achieved for $N=27$ - Fig.\,\ref{fig1}b. When $N=28$, the folded structure experiences a geometrical phase transition, resulting in a reduced curved region and forming two flat GNR-like strips. When $N\geq28$, the geometrical form becomes racket-like. Interestingly, the shape of the curvature is almost unchanged except the length of the flat regions becomes larger for larger value of $N$, as can be seen from Fig.\,\ref{fig1}b. The distance between the flat regions is similar to the interplane separation in graphite. It is ~$3.2$ {\it\AA}\ for even $N$ and ~$3.5$ {\it\AA}\ for odd $N$ number of Carbon lines. 

The type of open edges determines the stacking patterns of the folded ribbons. It is realized that $AA$ stacking is geometrically compatible with armchair GNRs, while $AB$ stacking is geometrically compatible with zigzag GNRs. Fig.\,\ref{fig1}c and Fig.\,\ref{fig1}d show the side and top views of the studied single folds. It turns out that all ribbons with odd $N$ take $AA$-stacking patterns. The folded ribbons with even $N$ occur in an orthorombic $AB'$ orientation characterized with a CC bond symmetrically situated above the center of each hexagon. We point out that $AB'$ stacking is highly unusual in regular graphite. It is one of the intermediate phases that may exist if very high pressure is applied \cite{PRL74-4015}, and yet here it occurs relatively easy by simple folding.

\begin{figure}[t]
  \includegraphics[scale=0.2]{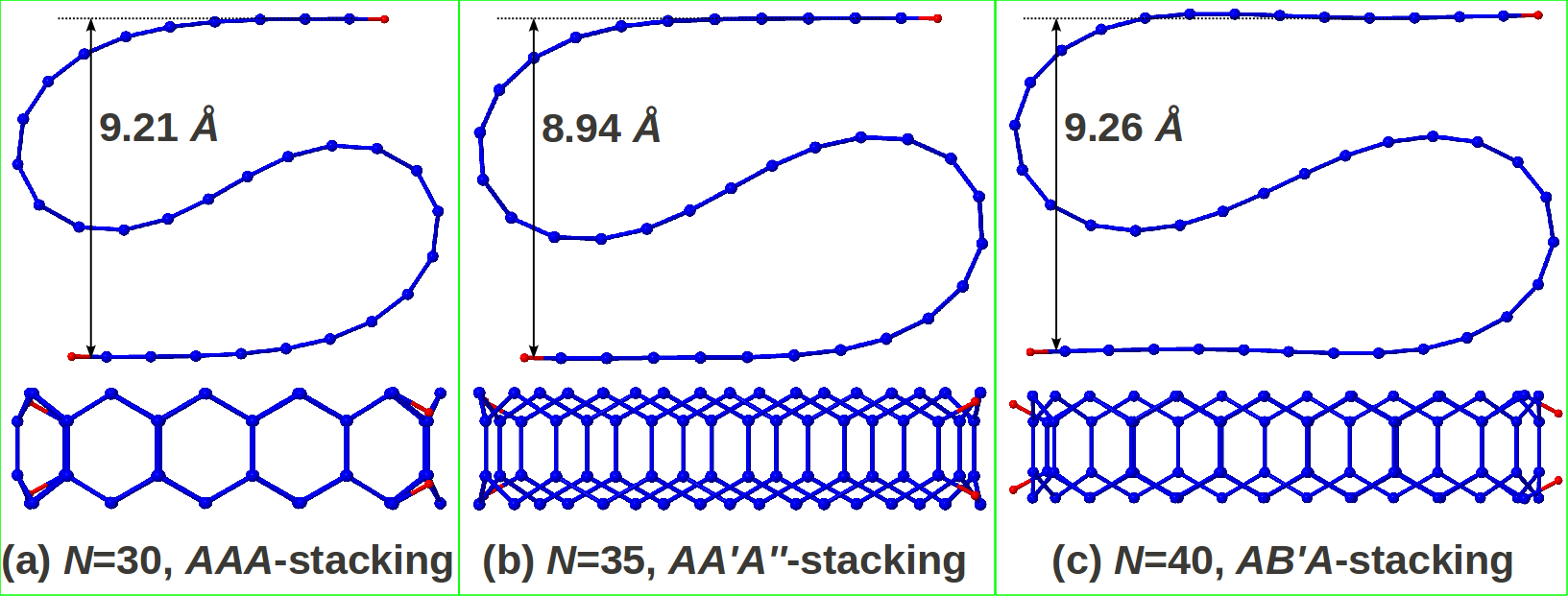}
  \caption{(Color Online) Double FGNRs for: (a) $N=30$ showing an $AAA$ stacking pattern; (b) $N=35$ showing an $AA'A''$ stacking pattern; (c) $N=40$ showing an $AB'A$ stacking pattern.}
  \label{fig2}
\end{figure}

Fig.\,\ref{fig2} shows some of the studied cases of the double folded structure. No pronounced geometrical phase transition is found here. However, as the GNR increases, the flat-like portions become larger, while the curved regions experience little change. The distance between the flat regions is always $\simeq9$ {\it\AA}, which is almost tripple the distance between the flat sides of the single folded ribbons with $N>28$. Considering the top view, the stackings for the different ribbons are resolved. It is obtained that they fall into categories distinguished by the number of carbon lines: for $N=3p$ ($p$ is integer), the pattern is $AAA$; for $N=3p+1$, the pattern is $AB'A$; and for $N=3p+2$, it is $AA'A''$. The last geometrical phase is described by having two CC bonds symmetrycally positioned in the Carbon hexagon (top view).

The graphene folding is a process that involves a balance between the elastic bending and the vdW interaction. The vdW attraction seeks to make a loop and it decreases the overall energy of the structure, while the bending tries to resist that tendency and it increases the total energy. The balance between these two effects determines the size and form of the loop as well as the flat-like regions. We find that the smallest ribbon for which single folding can occur is $N=17$ ($W=19.7\mbox{ {\it\AA}}$), while the smallest double folded ribbon has $N=28$ Carbon lines ($W=33.3\mbox{ {\it\AA}}$). Our calculations show that if the vdW potential is not taken into account, the single folded $N=17$ and double folded $N=28$ ribbons are not stable, indicating the importance of dispersion for the stability of the folds.

\begin{figure}[t]
  \includegraphics[scale=0.2]{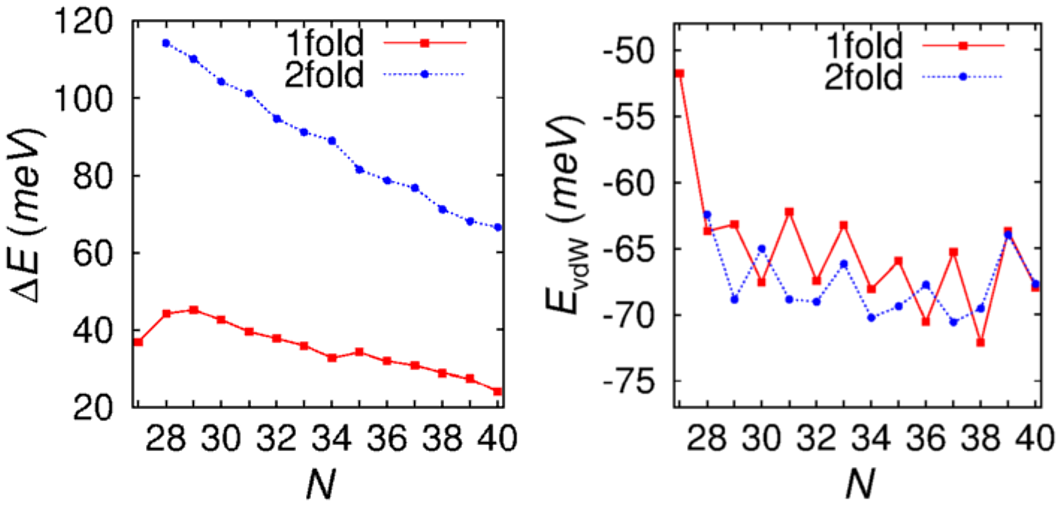}
  \caption{(Color Online) (a) Total energy per atom needed to create single (1fold) and double (2fold) FGNRs as a function of C atomic lines $N$; (b) vdW energy per atom for single and double FGNRs as a function of C atomic lines $N$.}
  \label{fig3}
\end{figure}

Further, we calculate the energy needed to achieve folding via the relation
\begin{equation}
 \Delta E_{1,2}=E_{1,2}^{(tot)}-E_0^{(tot)}.
\end{equation} 
where $E_{0,1,2}^{tot}$ are the total energies for the unfolded, single folded, and double folded GNRs, respectively. Results from the calculations are shown in Fig.\,\ref{fig3}. It is seen that in general, the energy decreases rather smoothly as the width becomes larger (larger $N$). We compare these results with the vdW energy obtained via the $D_2$ Grimme approach -  Fig.\,\ref{fig3}b. The vdW interaction is stronger when graphenes or GNRs are parallel and at a distance $\sim 3$ \AA. Thus $E_{vdW}$ becomes larger as the width of the ribbon increases reducing $\Delta E$. At the same time, the loop is smaller, thus less energy is needed for bending.

For the single FGNR with $N=27$, $E_{\mbox{\tiny vdW}}$ is smaller in magnitude as compared to the others, which reflects the lack of parallel flat-like portions since the geometrical phase transition occurs for $N=28$. It is interesting to note that in practically all of the other cases $|E_{\mbox{\tiny vdW}}|$ is in $62\div72\, eV$ range. The oscillatory-like behavior as a function of $N$ is attributted to the strong geometrical influence originating from the registry dependence of the C atoms for the different stackings. This effect has been previously realized in other graphitic nanostructures and it has also been attributed to the  geometrical dependence in such dispersive interactions \cite{JPCL1-1356}. One notes that for single FGNRs with even $N$ ($AB'$), $E_{\mbox{\tiny vdW}}$ is always lower than that of the closest odd $N$ ($AA$). The vdW energy for the double FGNR with $AB'A$ stacking also appears as local minima, however this trend is not pronounced as well.

\begin{figure}[t]
  \includegraphics[scale=0.2]{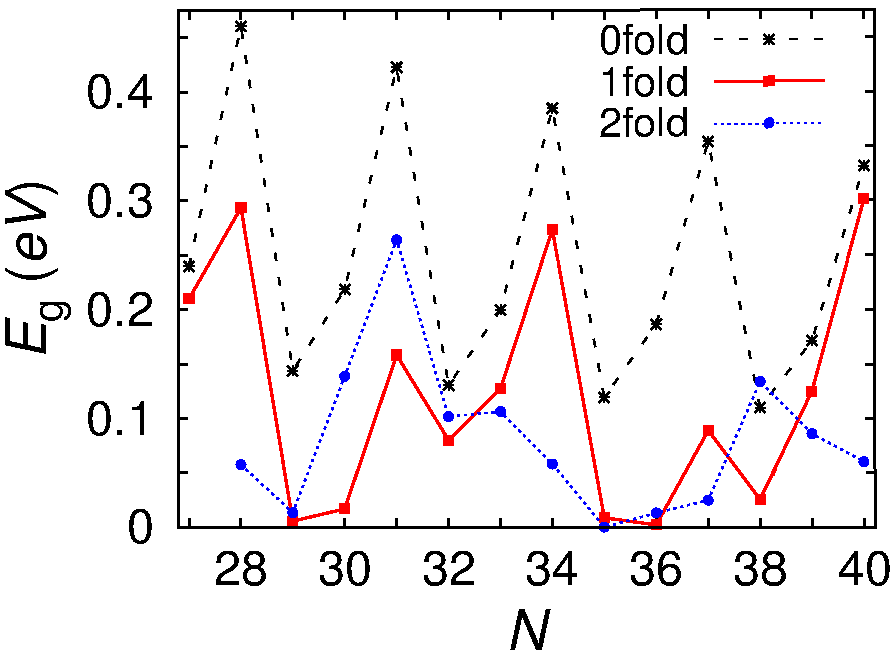}
  \caption{(Color Online) Energy band gap $E_g$ for unfolded (0fold), single folded and double folded GNRs as a function of C atomic lines $N$.}
  \label{fig4}
\end{figure}

\begin{figure}[t]
 \centering
 \includegraphics[width=0.48\textwidth]{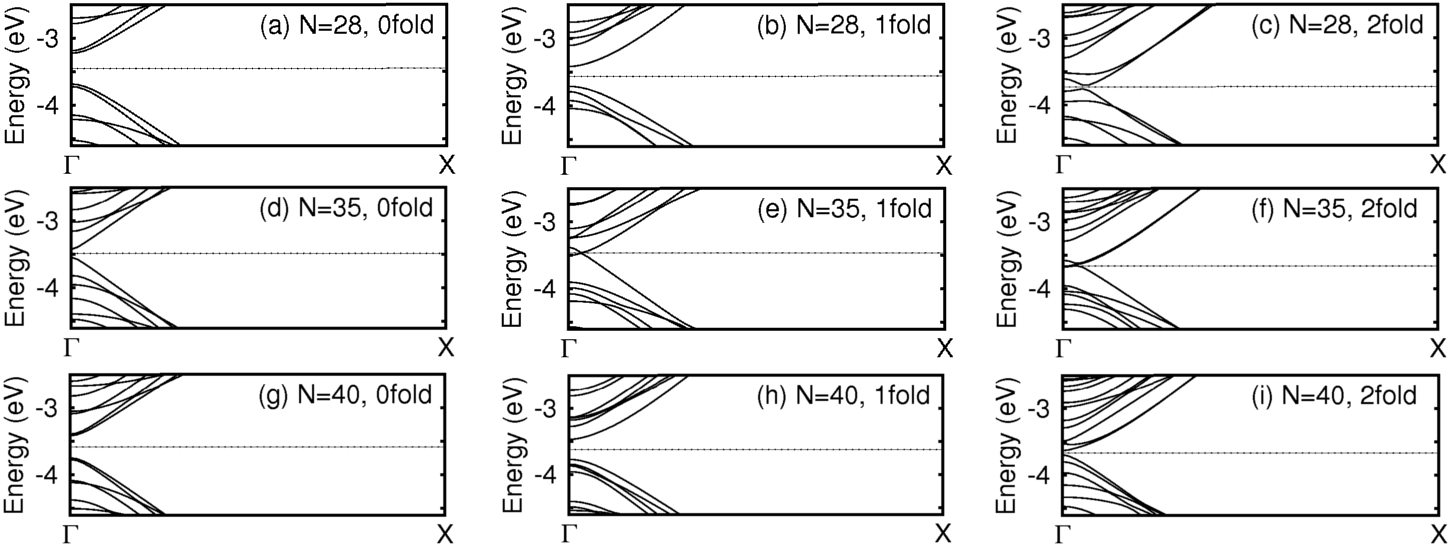}
 \caption{Energy bandstructures of unfolded, single and double FGNRs with $N=28$, $35$ and $40$ Carbon lines, as indicated in the top-right corner of each panel.}
 \label{fig5}
\end{figure}

The electronic structure of the different ribbons is also calculated within VASP. The resulting energy bandgaps are given in Fig.\,\ref{fig4}, while the energy bandstructure for three GNRs is shown in Fig.\,\ref{fig5}. All unfolded GNRs are found to be semiconductors which is directly related to the wave vector quantization along the finite width of the ribbon and the shorter H-C bond lengths at the ends. We find that the end H-C bonds are about $23\%$ shorter than the C-C bonds in the middle of the ribbon ($\sim 1.42 \AA$). This is particularly important for the $N=3p+2$ cases, which are predicted to be metals within a nearest neighbor tight binding model \cite{PRB83-245433}. If the tight binding model is modified to account for this difference in bond length, $E_g$ is non-zero in agreement with DFT calculations. The oscillatory and generally decreasing $E_g$ as a function of $N$ is in accordance with previously reported $ab$ $initio$ simulations of graphene ribbons \cite{NL6-2748}. The periodicity of 3 for these band gap oscillations are in the same classes as outlined above - $N=3p,3p+1,3p+2$. These are directly related to the $\pi$ nature of the ribbon orbitals having Fermi wavelength approximately four atomic sites along the width of the ribbon. Therefore, every time a C line is added, the Fermi wavelength changes in a 3-fold periodic pattern. Although the folded structures have reduced energy gaps as compared to the ones for the unfolded ribbons, the oscillatory-like behavior is preserved for the single FGNRs. Fig.\,\ref{fig4} shows that local maxima ($N=3p+1$) and minima ($3p+2$) in $E_g$ for single folds generally coincide with local maxima and minima for the unfolds. This behavior, however, is not preserved for the double folded structures. This is expected since additional stacking patterns occur in the 2fold GNRs.

The unfolded $N=28$ GNR is a semiconductor with relatively large bandgap $E_g=0.461\,eV$ at the $\Gamma$ point as seen in Fig.\,\ref{fig5}a. Folding the ribbon once results in moving the highest valence and lowest conduction bands closer - Fig.\,\ref{fig5}b. The energy bands of the double folded ribbon around the Fermi level are found to be almost touching away from the $\Gamma$ point, indicating that the system becomes semimetal. Similar behaviour is found for the single folded $N=35$ ribbon - Fig.\,\ref{fig5}e. The $N=40$ ribbon does not experience a semiconductor-semimetal transition due to the folding process. Folding it once and twice reduces the gap which is always at $\Gamma$ - Fig.\,\ref{fig5}g,h,i. Perhaps the most interesting transition is found for the double folded $N=35$ GNR. The system is a metal with a band crossing the Fermi level near $\Gamma$ - Fig.\,\ref{fig5}f.

Our calculations show that the vdW interaction is quite important for the magnitude of $E_g$ of the studied structures. If the vdW dispersion is not taken into account, the separation between the parallel portions is larger and consequently the coupling is weaker. For the $N=28$ 1fold GNR, for instance, the distance between the ends is $3.356 \AA$ without vdW and $3.156 \AA$ with vdW interaction, while for the $N=40$ 1fold GNR, the corresponding separations are $4.043 \AA$ and $3.568 \AA$. Comparing the energy gaps, we find that the vdW coupling reduces the band gaps. For example, for $N=39$ $E_g=0.138$ $eV$ (single fold) and $E_g=0.035$ $eV$ (double fold) without the vdW interaction, while $E_g=0.125$ $eV$ (single fold) and $E_g=0.0086$ $eV$ (double fold) when the vdW interaction is taken into account. Similarly, for $N=40$ $E_g=0.362$ $eV$ (single fold) and $E_g=0.246$ $eV$ (double fold) without vdW, while $E_g=0.301$ $eV$ (single fold) and $E_g=0.060$ $eV$ (double fold) with vdW interaction. 

Analyzing the electronic structure shows that the energy bands around $E_F$ are mainly determined by the $p$ orbitals from the edge atoms which are perpendicular to GNR axis. These can be considered as edge $\pi$ states. Upon folding, the interaction between the flat parallel-like portions lifts all degeneracies causing the energy bands to move closer. If the vdW dispersion is taken into account, the interlayer coupling is stronger leading to more distorted edge $\pi$ states, more prominent band splitting and thus smaller gaps. In addition, relatively small $\sigma-\pi$ hybridization from the orbitals located on the curved regions is found to contribute to the energy bands around $E_F$ due to the folding process. The smaller distance between the flat-like regions upon inclusion of the vdW dispersion in the simulations leads to slightly stronger bond changes at the curvature and the $\sigma-\pi$ hybridization is somewhat increased, although the dominant effect is still the interlayer interaction. The curved regions also have smaller influence on the electronic structure of larger ribbons. Their energy bandstructure is mainly determined by the parallel regions. The number of parallel layers further influences the bandstructure. If there are more such layers, the band splitting further increases due to the interlayer interaction and further reducing the band gap. Additional bands may also appear around $E_F$, which is the case in Fig. 5f, for example. Thus, in general, double folded GNRs have smaller $E_g$. For the double folded structures with smaller $N$, however, there are no significant flat regions since the widths of ribbons are relatively small. As a result, for several $N$ $E_g$ for 2fold is actually larger than $E_g$ for 1fold - Fig.\,\ref{fig4}.

These results illustrate how diverse the physical and energy band structures can be when effects due curvature, dispesive interlayer coupling, registry dependence, and structure size influence the system simultaneously. The stacking patterns and bandstructure characteristics functionalities are complex. The shift in registry can result in transforming an armchaired GNR from a semiconductor to another semiconductor, a semimetal, or even a metal. Folded ribbons with $AA$ and $AB'$ stackings can exist as semiconductors or semimetals depending on the number of C lines in each case. Taking into account the vdW dispersion is also important since it influences the energy band splitting and the resulting energy band gaps.

{\bf Conclusions.} Folded graphene nanoribbons with single and double closed edges were considered using first principles methods utilizing a DFT-D2 approach. We obtain geometrical phases of the studied structures in terms of various characteristic distances and stacking patterns, which are generally not found in open ended layered graphenes or GNRs. The energy band gaps and bandstructures are also obtained showing that the folding process together with relative carbon atomic positions can result in semiconductor-semimetal and semiconductor-metal transitions. The registry dependence of the dispersive interaction is calculated and it is found that it plays a decisive role for the stability and magnitude of the energy gaps.

This work was supported by the US Department of Energy under contract DE-FG02-06ER46297. We would like to acknowledge the use of the services provided by Research Computing at the University of South Florida. Communications with Prof. Xihong Peng about VASP and Dr. Tomas Bucko about the Grimme implemetation are recognized.

\end{document}